\documentclass[a4paper]{llncs}
   \usepackage{epsfig}
   \usepackage{color}
   \usepackage{latexsym}
   \usepackage{cite}
   \usepackage{graphicx}
   \usepackage{listings}
   \usepackage{multirow}
   \usepackage{tabularx}
   \usepackage{colortbl}
   \usepackage{amsmath}

\newenvironment{keywords}{
       \list{}{\advance\topsep by0.35cm\relax\small
       \leftmargin=1cm
       \labelwidth=0.35cm
       \listparindent=0.35cm
       \itemindent\listparindent
       \rightmargin\leftmargin}\item[\hskip\labelsep
                                     \bfseries Keywords:]}
     {\endlist}

\begin{document}

\pagestyle{headings}  

\title{Two Countermeasures Against Hardware Trojans Exploiting Non-Zero Aliasing Probability of BIST}
\titlerunning{Two Countermeasures Against Hardware Trojans}   
\author{Elena Dubrova\inst{1} \and Mats N\"aslund\inst{2} \and
Gunnar Carlsson\inst{3} \and John~Fornehed\inst{2} \and Ben Smeets\inst{2}}

\authorrunning{Elena Dubrova et al.}  
\institute{Royal Institute of Technology, Electrum 229, 164 40 Stockholm, Sweden\\
\email{dubrova@kth.se}
\and
Ericsson Research, Ericsson, F\"ar\"ogatan 6, 164 80 Stockholm, Sweden
\email{\{mats.naslund,ben.smeets\}@ericsson.com}
\and
Development Unit Radio, Ericsson, F\"ar\"ogatan 6, 164 80 Stockholm, Sweden
\email{gunnar.carlsson@ericsson.com@ericsson.com}
\and
BNET Systems \& Technology, Ericsson, F\"ar\"ogatan 6, 164 80 Stockholm, Sweden
\email{john.fornehed@ericsson.com@ericsson.com}
}

\maketitle

\begin{abstract}

The threat of hardware Trojans has been widely recognized by academia, industry, and government agencies. A Trojan can compromise security of a system in spite of cryptographic protection. The damage caused by a Trojan may not be limited to a business or reputation, but could have a severe impact on public safety, national economy, or national security. An extremely stealthy way of implementing hardware Trojans has been presented by Becker et al. at CHES'2012. Their work have shown that it is possible to inject a Trojan in a random number generator compliant with FIPS 140-2 and NIST SP800-90 standards by exploiting non-zero aliasing probability of Logic Built-In-Self-Test (LBIST). In this paper, we present two methods for modifying LBIST to prevent such an attack. The first method makes test patterns dependent on a configurable key which is programed into a chip after the manufacturing stage. The second method uses a remote test management system which can execute LBIST using a different set of test patterns at each test cycle. 
\end{abstract}


\section{Introduction} \label{intr}

Hardware is the root of trust of all secure computations and communications. Semiconductor chips are used in smart phones, cars, medical devices, etc., making the world around us more efficient, convenient and sustainable. Any cryptographic protocol, algorithm, or primitive is implemented either directly in hardware, or in software which is eventually run on hardware. This makes hardware an attractive target for adversaries. Attacks against critical infrastructure such as smart power grid, water supply systems, or traffic control may have catastrophic consequences for our society.

State-of-the-art integrated circuits are typically too complex and expensive to be designed and manufactured by one company. Instead, typically some companies focus on the design of integrated circuits, while other companies manufacture, test, package, personalize, and distribute the chips. Often a single integrated circuit is created using IP-blocks designed by other companies. With multiple players involved in the supply chain, there are plenty of opportunities to implant a {\em hardware Trojan} which opens a backdoor into a chip in spite of cryptographic protection~\cite{MiWW15}. Malicious modifications can be introduced into a design, for example, by tampering with a CAD-tool which is used for circuit's synthesis. The code of a CAD-tool is usually huge and it undergoes a continuous development. So, several extra lines which modify the original design to inject a Trojan may easily get unnoticed in a multi-million line code. Alternatively, a third-party-made IP-block might contain a backdoor that can be used to steal secret keys or extract internal chip data. Circuit modifications can also be made at the manufacturing stage, potentially affecting all chips or just some selected ones. Today's chips contain billions of transistors, so it is very difficult to identify which of them are not a part of the original design. Functional verification is further complicated by the fact that manufacturers are typically given a freedom to add redundant circuitry to a chip in order to increase manufacturing yield~\cite{GuP11}. 

The presence of hardware Trojans can be difficult to prove. For example, some PCs are claimed to contain malicious circuit modifications that allow a person who knows the modifications to remotely access a PC without the user's knowledge~\cite{lenovo}. However, it is still not confirmed if these claims are true or not. Some processors are suspected to contain backdoors deliberately implanted in their hardware Random Number Generators (RNGs) that make possible predicting RNG's output~\cite{intel_pr}. Again, it remains a conspiracy-theory story. 

We do not know if these stores are true or not. However, we cannot discount a possibility that such attacks may take place if they are feasible to implement.
For example, as demonstrated in~\cite{BeRPB13}, it is possible to reduce the security of a hardware RNG compliant with FIPS 140-2 and NIST SP800-90 standards from 128 to 32 bits by injecting stuck-at faults at the outputs of selected transistors. This can be done without disabling the Built-In Self-Test (BIST) logic which checks RNG's functionality at each power-up, without failing BIST tests, and without failing any randomness tests. 
Stuck-at faults can be injected a very stealthy way by modifying dopant types in the active region of transistors. Such dopant-level Trojans do not require adding any extra logic to the original design and therefore do not change its layout. As a result, the Trojan-injected circuit appears legitimate at all wiring layers. Even with the advanced imaging methods such as 
Scanning Electron Microscopy (SEM) or Focused Ion Beam (FIM), it extremely difficult to detect changes made to the dopant in a large design implemented with nanoscale technologies. To detect changes in dopant types, in addition to all metal layers, the contact layer has to be examined. High-quality imaging of the contact layer is significantly more costly than imaging of a metal layer~\cite{SuSFT14}. In addition, since only the dopants of a few transistors are modified, the change in the side-channel information is too small to be detected by side-channel analysis. Typically side-channel analysis can only detect sufficiently large Trojans that are at most three to four orders of magnitude smaller than the original design~\cite{AgBKRS07}. Trojans of a smaller size remain undetected.

The attack presented in~\cite{BeRPB13} exploits the fact that aliasing probability of BIST is non-zero due to the compaction of circuit's output responses. 
{\em Aliasing probability} is the probability that a fault-free circuit is not distinguished from a faulty one. If an $n$-bit compactor is used, the aliasing probability of BIST is $1/2^n$~\cite{DaOFE90}. In the traditional BIST, the same set of test patterns is applied to a circuit under test at each test cycle, and therefore the same compacted output response, called {\em signature}, is expected. 
Therefore, an adversary who knows the set of BIST test patterns can select suitable values for the Trojan that result in the same signature as a fault-free circuit signature. Since the aliasing probability is $1/2^n$, in order to inject a Trojan which does not trigger BIST, an adversary has to make $2^{n-1}$ simulation trials on average. The typical size of a BIST output response compactor
is 32 bits, so the attack is feasible in practice.

In this paper, we present two methods for modifying BIST to prevent such an attack. 
The paper is organized as follows.
Section~\ref{ht} reviews related work on hardware Trojans. 
Section~\ref{bist} gives a background on BIST.
Section~\ref{intel} describes the attack from~\cite{BeRPB13}.
Section~\ref{pres} presents two countermeasures against that attack.
Section~\ref{con} concludes the paper and discusses open problems.

\section{Background and Related Work} \label{ht}

In this section, we give a brief introduction to hardware Trojans, describe previous approaches for their detection and prevention, and analyze if these approaches can be used to combat the dopant-level Trojans from~\cite{BeRPB13}.  

\subsection{Definition}

A hardware {\em Trojan} is a malicious change of a design that makes it possible to  
bypass or disable the security of a system~\cite{TeK10}.
The purpose of Trojan insertion can be either to leak confidential information to the adversary, or to disable/destroy a chip. 

Hardware Trojans has been known for a while (also referred to as {\em sleeper cells}~\cite{ht14}), but previously it was very difficult to inject a Trojan into the supply chain. In today's globalized world where the manufacturing is outsourced and the use of third-party IP from small and relatively new vendors is widespread, this is no longer a problem.

There are two different kinds of Trojans~\cite{TeK10}. 
{\em Functional} Trojans add or remove transistors, gates or other components to/from the original design.
{\em Parametric} Trojans reduce the reliability of a chip by thinning of wires, weakening of transistors, or subjecting the chip to radiation. A chip with a parametric Trojan produces errors or fails every time the affected component is loaded intensely. 

\subsection{Detection of Hardware Trojans}

Hardware Trojan detection methods can be divided into {\em pre-manufacturing} and {\em post-manufacturing}~\cite{BhH14}.

Pre-manufacturing methods~\cite{PoNNM09,Oy15,Ca15} aim at detecting Trojans inserted into a HDL description of the design, or an RT- or gate-level netlist during the design stage. Such malicious modifications can be done, for example, by an untrusted employee or an intentionally modified CAD-tool. 
Post-manufacturing methods~\cite{BeRPB13,SuSFT14} focus at detecting Trojans added during or after the manufacturing stage. Such Trojans can be inserted, for example, by an untrustworthy manufacturer or a third party performing packaging or personalization.

Current techniques for Trojan detection include:
\begin{enumerate}
\item {\em Visual inspection} in which the layers of a chip are repeatedly removed and the exposed circuitry is scanned
using various high-resolution imaging methods~\cite{Sk11,SuSFT14,Co15}; 
\item {\em Side-channel analysis} in which signals emitted by a chip, e.g. power, path delays, or electromagnetic radiation  are measured~\cite{Ji08,YiM08,RaPT10,Ngo15};
\item {\em Testing} (ATPG or BIST), in which test stimuli are applied to a chip and its output is monitored to detect functional differences from the specification~\cite{DuNS14,DupBBR15}.
\end{enumerate}

\paragraph {\bf Visual inspection}

To perform a visual inspection, a target chip is first depackaged. Then, the layers of the chip are removed one-by-one by polishing or etching and, for each exposed layer, images are taken with a high-resolution imaging method such as Scanning Electron Microscopy (SEM) or Focused Ion Beam (FIM)~\cite{Sk11}. These images are then compared to the corresponding images of a golden chip to detect possible differences. A {\em golden} chip is a chip which is known to have the correct functionality complying with the specification, without any malicious modifications. 

Changes in metal wires and transistors can typically be detected with a high probability. It is more difficult to detect changes made to the dopant, especially in nanoscale technologies. To detect changes in dopant types, in addition to all metal layers of a chip, the contact layer has to be examined. Imaging of a contact layer requires a higher magnification compared to the one of metal layers. According to~\cite{SuSFT14} high-quality imaging of the contact layer is 16 times more costly than imaging of a metal layer.

As we can see, visual inspection is expensive and time consuming process. It also destroys the inspected chip. Therefore, it can only be applied to a small number of chips. So, although in theory it is possible to detect dopant-level Trojans from~\cite{BeRPB13} using advanced imaging methods such as SEM or FIM, on practice it is extremely difficult, especially for large designs implemented with state-of-the-art technologies.

\paragraph {\bf Side-channel analysis}

In side-channel analysis, signals emitted by the examined chip, e.g. power~\cite{RaPT10}, path delays~\cite{Ji08,YiM08}, or electromagnetic radiation~\cite{Ngo15}, are measured and compared to the corresponding signals of a golden chip. Typically side-channel analysis can only detect Trojans 
that are at most three to four orders of magnitude smaller than the original design~\cite{AgBKRS07}.

The dopant-level Trojans from~\cite{BeRPB13} modify only a small number of transistors (4 out of the 32)
in a small number of flip-flops in a design. The dominant part of the gates remains unchanged and therefore behaves according to the specification. As a result, the change is the side-channel information is too small to be detected by side-channel analysis. 

Another technique to detect Trojans, proposed in~\cite{RaJK11}, is to add to a design 
redundant Trojan detection circuitry that can decide if the design was modified during the manufacturing.
For example, gates that transform some parts of the design into ring oscillators can be added.
During the analysis, frequencies of these ring oscillators can be compared to the corresponding frequencies 
of a golden chip to detect if the design was changed. A problem with such a method, as well as with any other method based on extra Trojan detection circuitry, is that the circuitry itself can be maliciously modified at the manufacturing stage not to trigger the Trojan.

\paragraph {\bf Testing}

For a design is using the traditional Design-for-Test (DFT) architecture with a JTAG port and scan chains, it is possible to access the internal content after the manufacturing and thoroughly check the functionality using various external testing methods~\cite{atpg}. Note, however, that test sets are generated for the original, Trojan-free netlist and typically for a specific fault model (stuck-at, transition, small-delay defects, etc). Therefore, they may not be able to detect Trojans as efficiently as they detect manufacturing defects.

For a design without JTAG and scan scan, BIST can used. However, as we mentioned in Section~\ref{intr}, it is possible to exploit the non-zero aliasing probability of BIST and make Trojan modifications so that the Trojan-injected circuit produces the same expected signature as a fault-free circuit. For this reason, the dopant-level Trojans from~\cite{BeRPB13} are resistant to the conventional BIST.

\subsection{Prevention of Hardware Trojans}

Some countermeasures have also been developed to protect against activation of certain Trojans,
or to maintain secure operation in presence of unknown Trojans, see~\cite{BeHN11} for an excellent overview. 

The former methods typically involve utilizing data guards such as scrambling or obfuscation, or hardening the architecture against specific triggers. For example, control circuitry that makes Trojan activation difficult can be added between the untrusted IPs~\cite{Wa11}.

The latter methods are usually implemented by replication, fragmentation and voting, as in the traditional fault-tolerant design~\cite{Du13book}. For example, a part of hardware design which is not covered by simulation/verification can be covered with software~\cite{HiKM10}. 

Such prevention methods typically require the addition of a large amount of redundancy to the original design. Therefore, they are not suitable for protecting devices with constrained resources.

\section{Conventional Built-In Self-Test} \label{bist}

BIST was introduced in the 80's with the purpose of combating the raising complexity of external testing~\cite{McC85}. In BIST test generation and response capture logic are incorporated on-chip. On-chip circuitry usually works at a much higher frequency than
an external tester. So, by embedding the test pattern generator on chip, test application time can be reduced. By embedding the output response analyzer on chip, time to compute the circuit
response can be reduced as well.

There are different types of BIST. Logic BIST (LBIST), on which we focus in this paper, is used for testing random digital logic~\cite{RaT98}. Memory BIST (MBIST) is designed for testing memories~\cite{Sh02}.

Cryptographic modules such as random number generators, block or stream ciphers, cryptographic hash functions, etc. often employ LBIST to make possible periodic fault detection of functional circuits during their lifetime. A random hardware fault can compromise the security of a system. For example, if the output of a pseudo-random number generator used in a stream cipher gets stuck to 0, then the stream cipher will be sending messages unencrypted.

\begin{figure}[t!]
\begin{center}
\includegraphics[width=0.8\columnwidth]{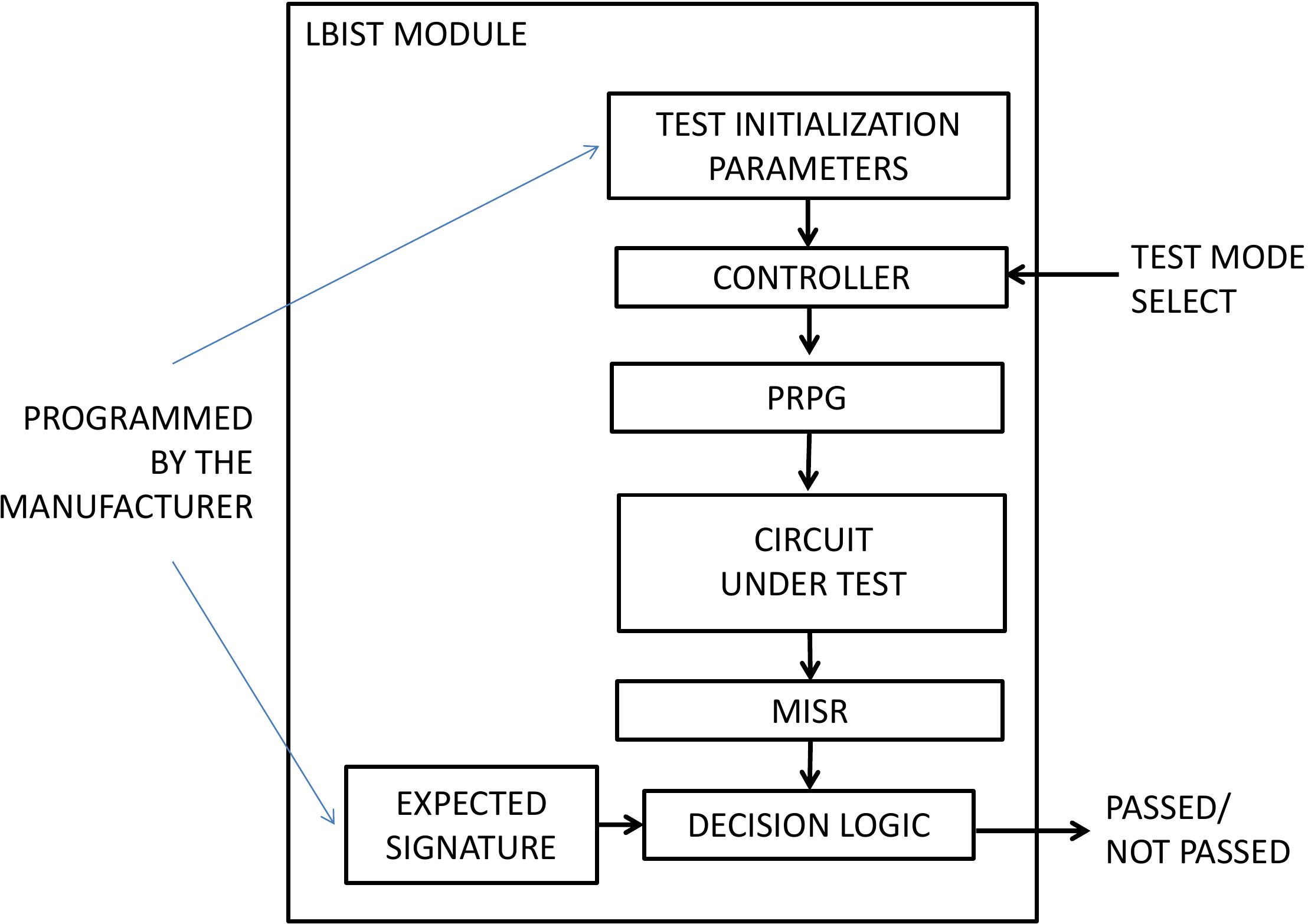}
\caption{Traditional LBIST.}\label{f1} 
\end{center}
\end{figure}

The traditional LBIST employs a Pseudo-Random Pattern Generator (PRPG) 
to generate pseudo-random test patterns that are applied to the circuit under test and an output response compactor for obtaining the cumulative value of the output responses of the circuit to these test patterns, called {\em signature} (see Figure~\ref{f1})~\cite{LiCDP15}. Faults are detected by comparing the computed signature to the expected "good" signature.

Theoretically, it is possible to generate a complete set of test patterns off-line using some Automatic Test
Pattern Generation (ATPG) method~\cite{atpg} and store this test set in an on-chip Read Only Memory (ROM). However, such a
scheme does not reduce the cost of test pattern generation and requires a very large ROM. Several gigabits of
test data may be required for a multi-million gate design~\cite{HeF99}. 
Instead, pseudo-random patterns generated by
a Linear Feedback Shift Register (LFSR) are usually used as test patterns~\cite{McMMW88}. 
LFSRs are simple, fast, and easy to implement in hardware~\cite{Golomb_book}.

The output response compactor is usually implemented by a  
Multiple Input Signature Register  (MISR). Since the output response is compacted, a faulty circuit may produce the same signature as the correct circuit. This is known as an {\em aliasing} error. If an MISR with a primitive generator polynomial is used\footnote{An irreducible polynomial of degree $n$ is called {\em primitive} if the smallest $m$ for which it divides $x^m+1$ is equal to $2^n-1$\cite{Golomb_book}.}, then the aliasing probability is bounded by $1/2^n$~\cite{DaOFE90}, where $n$ is the size of the MISR. 

LBIST controller contains control circuitry that administrates the LBIST testing process: generation of pseudo-random test patterns, their application to the circuit under test and compaction of responses of the circuit to these test patterns. In operation, the controller initializes the PRPG with the initial state defined by the test initialization parameters. After the initialization, controller counts the number of test patterns generated by the  
PRPG and stops the PRPG when a pre-defined number of patterns is generated. 

Pseudo-random patterns generated by the PRPG are applied to the circuit under test and propagated through its components. The resulting responses are fed into the MISR. The MISR computes the signature and forwards it to the decision logic.

Decision logic compares the signature computed by the MISR to the expected signature and makes a decision whether the circuit under test passed or failed the test cycle. If the MISR signature matches the expected signature, the circuit passes the test cycle; otherwise it fails the test.  

The test initialization parameters and the expected signature are typically stored in a memory or hard-wired during the manufacturing stage. LBIST is usually performed automatically at power-up and restart, or in response to an external trigger, e.g., if a hardware or software supervising the chip indicates a fault. In addition, LBIST may be initiated by an operator, e.g., for debugging purposes when a faulty chip is sent for repair.

\section{Attack of a Hardware Random Number Generator} \label{intel}

In this section, we describe the attack on a hardware RNG presented in~\cite{BeRPB13}. 

The RNG considered in~\cite{BeRPB13} consists of an entropy source and a digital post-processing unit~\cite{intel}.
The digital post-processing unit contains an Online Health Test (OHT) module and a cryptographically secure Digital Random Bit Generator (DRBG). The OHT monitors the random numbers from the entropy source to guarantee that they have a required minimum entropy. DRBG includes a conditioner and a rate matcher. The conditioner computes new seeds for the rate matcher. On the based of these seeds, the rate matcher computes 128-bit random numbers. This is done  by performing an Advanced Encryption Standard~\cite{aes} (AES)-based encryption with a 128-bit seed $c$ and a 128-bit encryption key $K$. 

The generated random numbers are tested against a range of statistical tests in order to be NIST SP800-90 and FIPS 140-2 compliant.
In addition, to be compliant with FIPS 140-2, the RNG contains an LBIST module which checks the functionality of the RNG at each power-up. When LBIST is initiated, the entropy source is disconnected and replaced by a 32-bit LFSR which generates pseudo-random test patterns. The 32-bit MISR compacts the resulting output responses of the rate matcher into a signature.
This signature then is compared to a hard-wired expected signature to detect faults in the conditioner and the rate matcher. If two signatures are the same, the RNG passes the LBIST.


In the Trojan-free case, the probability that an adversary successfully guesses a random
number generated by the RBG is $1/2^{128}$, i.e. the attack complexity is 128-bits. 
It was shown in~\cite{BeRPB13} that this complexity can be reduced to $n$ bits by
fixing to constants all flip-flops that store the key $K$ and all but $n$
flip-flops that store the seed $c$. As a result of these modifications, a 128-bit random number generated by the rate matcher
depends on $n$ random bits and $256-n$ constant bits known to the adversary.
Therefore, the adversary can predict 128-bit random numbers with the
probability of success $1/2^n$. On the other hand, if $n$ is sufficiently large, the Trojan RNG will still pass all the statistical tests, because by the output generated by the AES generates has very good statistical properties, even if its inputs differ in a few bits only. It was shown in~\cite{BeRPB13} that the Trojan-injected RNG passed all randomness tests for the case of $n = 32$. 

In order to pass LBIST check as well, the adversary has to find constant values for the 128 bits of the key $K$ and the $128-n$ bits of the seed $c$ which produce the same signature as a fault-free circuit. This can be done by simulating the circuit with different stack-at faults injected into the flip-flops that store $K$ and $c$. If $n = 32$, then the adversary needs to do $2^{31}$ simulation trials on average to succeed. In Section~\ref{pres1}, we show an example of such an attack. 

To summarize, the attack presented in~\cite{BeRPB13} has shown that non-zero aliasing probability of LBIST can be advantageously exploited to inject hardware Trojans in an RNG which is compliant with FIPS 140-2 and NIST SP800-90 standards. In the next section, we show how LBIST can be modified to prevent such attacks.

\section{Proposed Countermeasures Against Trojans} \label{pres}

In this section, we present two countermeasures against the hardware Trojans exploiting non-zero aliasing probability of LBIST.
Both countermeasures are based on the observation that, in order to prevent such attacks, it is sufficient to make the LBIST test patterns unknown until the manufacturing stage is completed.

\subsection{Keyed LBIST} \label{pres1}

\begin{figure}[t!]
\begin{center}
\includegraphics[width=0.8\columnwidth]{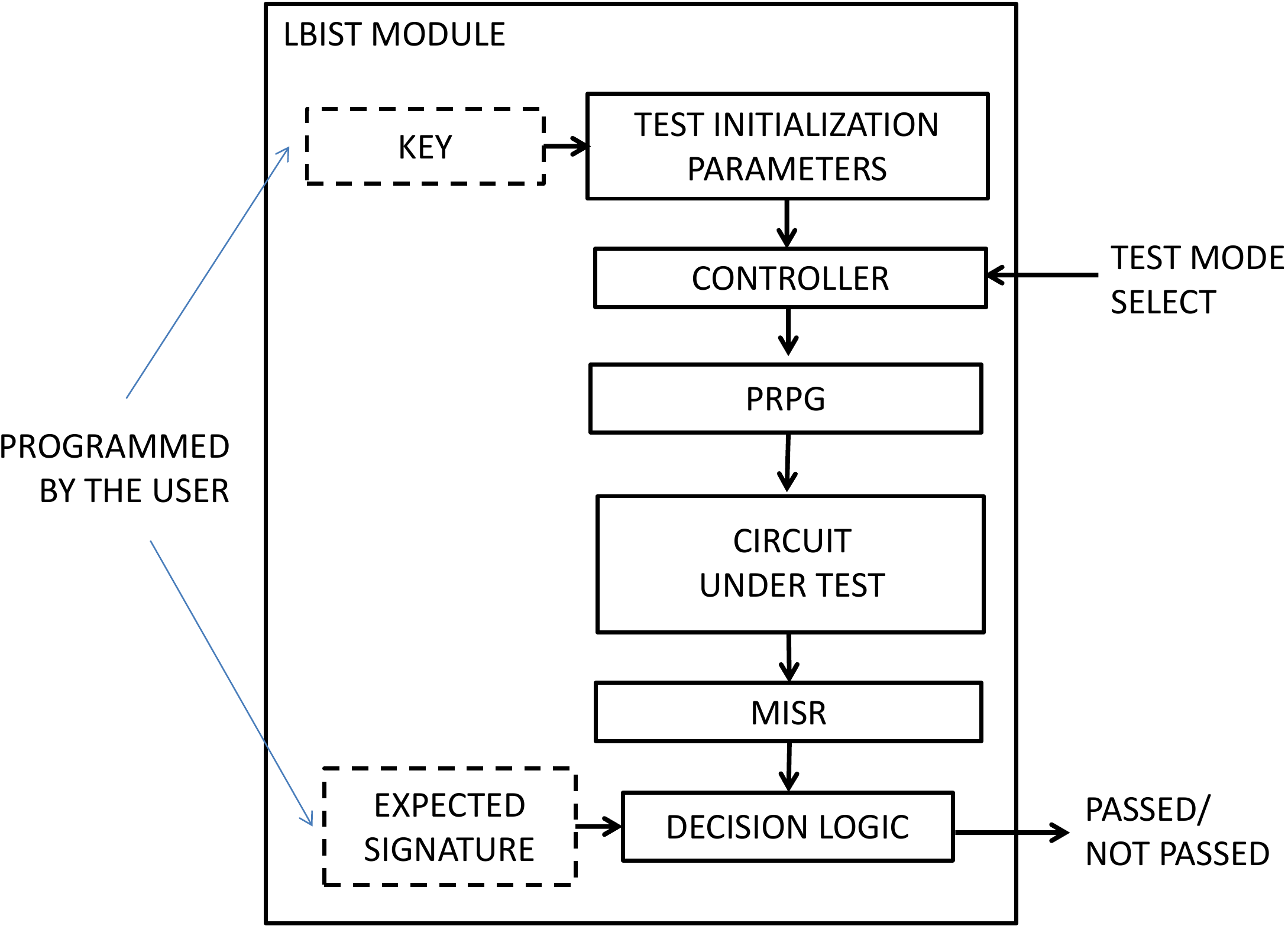}
\caption{The presented keyed LBIST.}\label{f2} 
\end{center}
\end{figure}

One possibility to mitigate the dopant-level Trojans presented in~\cite{BeRPB13} is to make the initial state of the PRPG dependant on a configurable {\em key}, as shown in Figure~\ref{f2}. The PRPG needs to be adapted to generate test patterns based on an initialization value which is derived from the key.

To eliminate the possibility that the key is leaked to the manufacturer by an untrusted employee, the key should be decided and programmed into the chip {\em after} the manufacturing stage. 
Once the key is decided, the expected MISR signature can be computed by simulation and programmed into the chip as well.
The programming of the key and the signature can be done either by the user or by trusted third party which performs chip personalization.

The key and the signature can be stored into an on-chip non-volatile memory, such as Flash or Electrically Erasable Programmable Read-Only Memory (EEPROM), or by means of programmable fuses. 
Almost all modern non-trivial integrated circuits contain a fusebox which is used e.g. for repair of on-chip memories at power-up or for storing secret keys. 
Even if a chip does not contain a JTAG port,
a fusebox or an on-chip non-volatile memory can be accessed for programming "from inside" by a dedicated on-chip software. The software is typically controlled by a PC connected to a board on which the chip is mounted though a board connector. 

Note that the key does not have to be kept secret after the chip has been fabricated. If the currently stored key and the signature become compromised, e.g., an adversary gains knowledge of them when the chip is sent for repair or maintenance, a new key and a new signature can be
programmed by the user upon receiving the chip back. By re-programming the signature, we will also be able to detect counterfeit in the case the original chip has been replaced during repair or maintenance.

\begin{figure}[t!]
\begin{center}
\includegraphics[width=0.45\columnwidth]{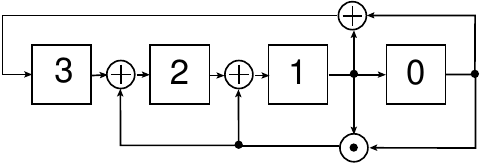}
\caption{4-bit NLFSR from the example.}\label{f4} 
\end{center}
\end{figure}

\begin{table}[t!]
\begin{center}
\begin{tabular}{|c||c|c||c|c|}  \hline     
Test    & \multicolumn{2}{|c||}{Fault-free case} & \multicolumn{2}{c|}{With fault injected} \\ \cline{2-5}
pattern    & NLFSR         & MISR          & NLFSR         & MISR         \\ 
from LFSR    & response        & signature         & response        & signature         \\ 
$(x_3 x_2 x_1 x_0)$ & $(x_3 x_2 x_1 x_0)$ & $(x_3 x_2 x_1 x_0)$ & $(x_3 x_2 x_1 x_0)$ & $(x_3 x_2 x_1 x_0)$  \\
\hline
1011  & 0011  & 0011  & 0001  & 0001  \\
1010  & 1101  & 0101  & 1101  & 0100  \\
0101  & 1010  & 0001  & 1000  & 1010  \\
1101  & 1110  & 0111  & 1100  & 1001  \\
1001  & 1100  & 0110  & 1100  & 0001  \\
1011  & 0011  & 0000  & 0001  & 1000  \\
1010  & 1101  & 1101  & 1101  & 1001  \\
0101  & 1010  &  {\cellcolor[gray]{0.9}} 0101 & 0001  & \cellcolor[gray]{0.9} 0101  \\ \hline
\end{tabular}
\end{center}
\caption{Example of a hardware Trojan not detected by LBIST.} \label{t1}
\end{table}

As an example, consider an RNG implemented by a 4-stage Non-Linear Feedback Shift Register (NLFSR)  shown in Figure~\ref{f4}. This NLFSR generates pseudo-random numbers in the range $\{1,2,\ldots, 15\}$. The sequence of generated numbers is defined by the following feedback functions:
\[
\begin{array}{l}
f_0(x_1) = x_1 \\
f_1(x_0,x_1,x_2) = x_2 \oplus x_0 x_1 \\
f_2(x_0,x_1,x_3) = x_3 \oplus x_0 x_1 \\
f_3(x_0,x_1) = x_0 \oplus x_1.
\end{array}
\]
where $x_i$ represents for the state variable of the $i$th stage of the NLFSR, $i \in \{0,1,2,3\}$,
$f_i$ is the feedback function of the $i$th stage and "$\oplus$" is the XOR. 
At each clock cycle, 
the next state of the NLFSR is computed from its current state 
by simultaneously updating the value of each stage $i$
to the value of the corresponding feedback function $f_i$~\cite{Du09j}.

Suppose that the LBIST which is used to test such an RNG uses a 4-bit LFSR with the connection polynomial $1 \oplus x \oplus x^2 \oplus x^3 \oplus x^4$ as a PRNG and a 4-bit MISR with the connection polynomial $1 \oplus x^3 \oplus x^4$  as an output compactor. 
We assume that, at each clock cycle, the current 4-bit state vector of the LFSR is used as a test pattern. The LFSR is initialized to some non-zero state and the MISR is initialized to (0000).
Both, LFSR and MISR are implemented in the Galois configuration.

The probability that an attacker successfully 
guesses the number generated by an $n$-bit NLFSR is $1/2^n$.
However, by setting $k$ internal flip-flops of the NLFSR to a constant value
it is possible to reduce the complexity of the attack to $1/2^{n-k}$. 

Suppose that the attacker knows that the initial state of the LFSR is (1011) and that 8 test patterns are applied to compute the MISR signature. Then, the attacker can calculate the expected MISR signature by simulation. From the 3rd column of Table~\ref{t1} we can see that this signature is (0101). The attacker can search which of the NLFSR's flip-flops should be fixed to constant-0 or constant-1 value in order to get the same signature. In our example, the signature (0101) can be obtained by fixing the flip-flop corresponding to the stage 1 of the NLFSR to 0 (see the last column of Table~\ref{t1}). So, the attacker can inject such a fault and reduce the complexity of the attack by one half. 

The presented method mitigates this problem because it makes the expected signature unknown before the circuit is manufactured. Therefore, an adversary who wants to inject a Trojan at the manufacturing stage does not know how to modify the circuit to get the same signature as the fault-free circuit signature.

\subsection{Remotely Managed LBIST} 

\begin{figure*}[t!]
\begin{center}
\includegraphics[width=0.9\columnwidth]{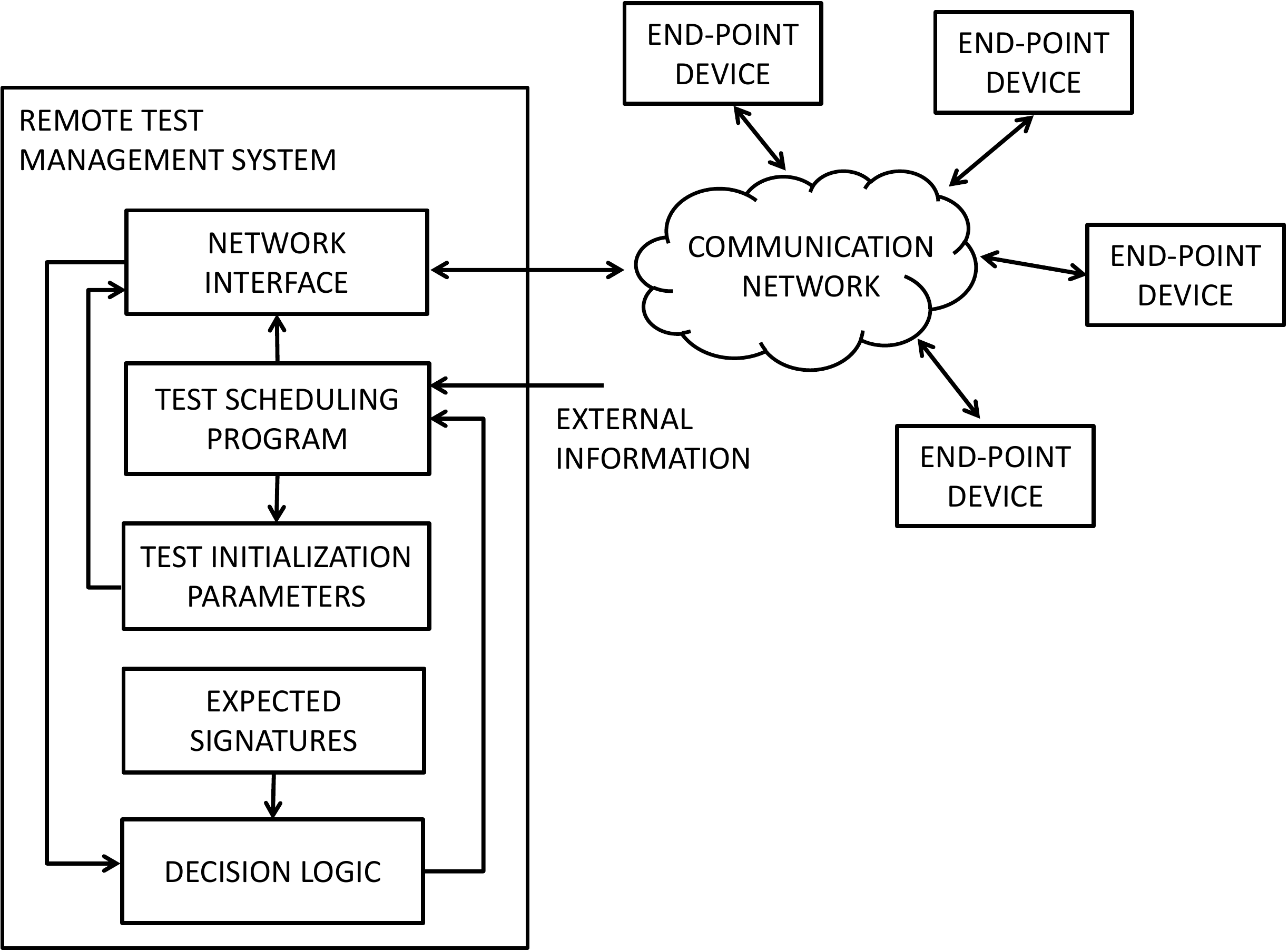}
\caption{Remote test management}\label{f5} 
\end{center}
\end{figure*}

Another way to mitigate the dopant-level Trojans from~\cite{BeRPB13} is to modify 
LBIST so that it uses a different set of test patterns at each test cycle. 
We propose to implement it by using a centralized
remote test management system which monitors all end-point devices in the same network, as shown in Figure~\ref{f5}. The test management system is expected to have sufficient storage and/or computational resources to either pre-compute and store the expected signatures for a given set of test initialization parameters, or to compute them on-the-fly by simulation. The use of test management system allows us to remove some of the LBIST functionality (test initialization parameters, expected signatures, and decision logic) from devices under test. It also does not require the addition of keys, as in the previous countermeasure. However, the price we pay is extra communication load. Depending on the application requirements, the former might be preferable to the latter, or vice versa.

The remote test management system contains a test scheduling program, test initialization parameters, expected signatures, decision logic and network interface. Upon deciding to initiate a test cycle for all or some devices, the test scheduling program 
instructs the test initialization parameters module which parameters to send. These parameters are transmitted through the network interface and the communication network to the selected end-point devices.

\begin{figure}[t!]
\begin{center}
\includegraphics[width=0.7\columnwidth]{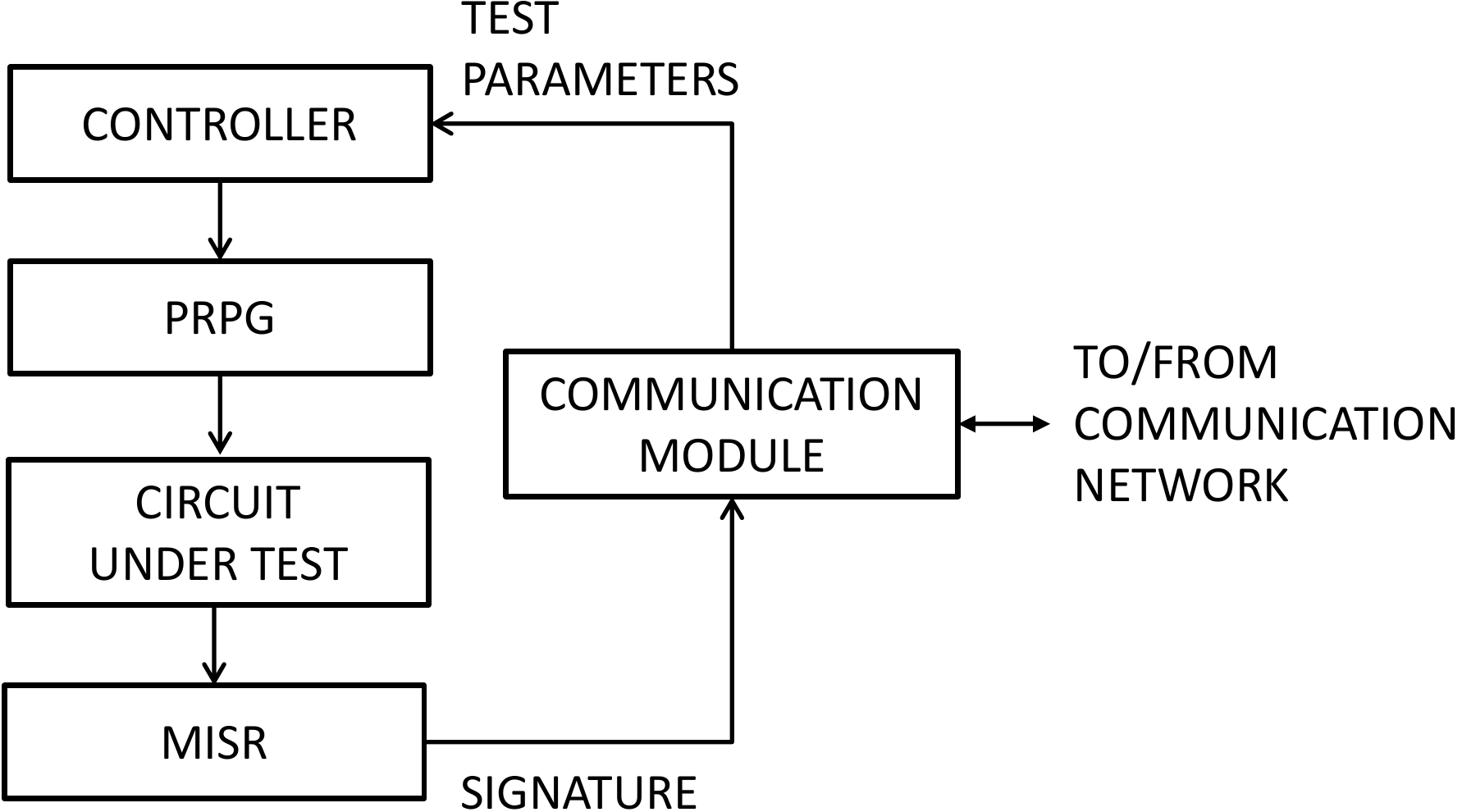}
\caption{Remotely managed LBIST module.}\label{f6} 
\end{center}
\end{figure}

The end-point devices contain a remotely managed LBIST module whose diagram is shown in Figure~\ref{f6}. We assume that the board implementing a device contains a communication module, e.g. a modem, which performs the communication between the device and the remote test management system, an CPU which runs processes related to the communication, and Basic Input/Output System (BIOS) which provides the functionality and interface for communication during LBIST.

Then the test initialization parameters are received by the devices, the LBIST test cycle proceeds as normally and the signature is computed. This signature is returned to the remote test management system for the analysis on passing/not passing the test.

The test management system takes "human-like" decisions regarding which devices to test, when, and how. These decisions can be taken based on the information received  
by monitoring global external factors such as environment, the interaction between devices, abnormal responses, etc. For example, meteorological sensors that register the wind in various locations within a given area may be tested immediately after harsh weather conditions, e.g a thunderstorm. As another example, a device may request the management system to test another device if several attempts to communicate with it failed. In both examples, faults might be detected earlier, implying higher availability and safety.
The idea of context-aware automation and decision optimization is not new~\cite{WaD09}, but to our best knowledge it has not been applied to LBIST before.  

Note that, by executing LBIST using a different set of test patterns at each cycle, the presented method is able not only to detect unanticipated random and malicious faults, but also to cover different subsets of faults. Therefore, it has a potential to provide a higher fault coverage compared to the traditional LBIST. This is important for applications requiring high reliability.

In the scenario we described above, the remote test management system sends to a device the test initialization parameters and the device replies with a signature.  
Another scenario is possible, in which the remote test management system sends to a device both, the test initialization parameters and the expected signature. Then, the device computes the signature, compares it to the received expected signature and replies with pass/not passed. While such a case would involve the same volume of data transferred, it
might be preferable for applications in which the downlink bitrate of the receiving device is higher than its uplink bitrate.

\section{Conclusion} \label{con} 

We introduced two countermeasures against the hardware Trojans exploiting non-zero aliasing probability of LBIST. Both countermeasures are based on the observation that, in order to prevent such attacks, it is sufficient to make the test patterns unknown until the manufacturing stage is completed. The first countermeasure uses a configurable key which is decided and programed into a chip after its manufacturing. The test patterns are made dependent on the key. 
The second countermeasure transfers some of the LBIST functionality from the device under test to a remote test management system which has sufficient computational resources to execute LBIST using a different set of test patterns at each test cycle. 
Depending on the application requirements, the former approach might be preferable to the latter, or vice versa.

There is no "silver bullet" method that can protect against all possible types of hardware Trojans or other adversarial attacks. In parallel with new countermeasures, more complex attacks are being developed. 
Moreover, we are dealing with a two-ended stick - a method originally designed as a countermeasure can be later turned into an attack, and vice versa.  
For example, advanced visual inspection methods for Trojan detection can be used by IP thefts to reverse-engineer chips.
Similarly, if side-channel analysis techniques for detecting Trojans that affect only a tiny fraction of a design are invented, they are likely to give rise to more effective side-channel attacks.

Future work remains investigating if there are attacks which can go around the presented countermeasures and what can be done to avoid them.

\section*{Acknowledgement}
The first author was supported in part by the research grant No SM14-0016 from the Swedish Foundation for Strategic Research.
 

\end{document}